\documentclass[
aps,
prd,
preprint,
11pt,
longbibliography,
superscriptaddress,
nofootinbib,
floatfix
]{revtex4-1}
\usepackage{amsmath,amssymb,amsfonts,bm}
\usepackage{graphicx}
\usepackage{booktabs}
\usepackage{xcolor}
\usepackage[
colorlinks=true,
linkcolor=blue,
citecolor=blue,
urlcolor=blue
]{hyperref}

\begin{document}

\title{Decoherence Effects on Primordial Black Holes and Scalar-Induced Gravitational Waves}

\author{Waqas Ahmed}
\thanks{E-mail: \href{mailto:waqasmit@hbpu.edu.cn}{waqasmit@hbpu.edu.cn}}

\affiliation{
Center for Fundamental Physics and School of Artificial Intelligence,
Hubei Polytechnic University, Huangshi, China
}


\begin{abstract}
Primordial black holes (PBHs) form when large primordial curvature
perturbations re-enter the Hubble radius and exceed the classical collapse
threshold.  These perturbations originate as quantum fluctuations of the
inflationary vacuum, motivating a quantum-information description of the
PBH-producing scalar sector.  We develop a conservative extension of the
standard PBH and scalar-induced gravitational-wave (SIGW) framework in which
Gaussian quantum discord is used as a diagnostic of residual quantum
correlations, not as a new PBH-formation criterion.  We describe each
\((\bm k,-\bm k)\) pair as a two-mode Gaussian state and show that, in the pure
squeezed limit, discord grows rapidly with the squeezing parameter, so low
discord thresholds are automatically satisfied for strongly squeezed modes.
The nontrivial regime is the mixed state produced by decoherence.  Using a
Lindblad open-system description, we motivate a Gaussian loss channel for the
scalar covariance matrix and distinguish the discord from a covariance-survival
factor \(Q_{\rm dec}(k)\).  If the decoherence channel suppresses the scalar
two-point covariance, PBH abundance can be affected through the classical
collapse variance, while the SIGW spectrum is modified more directly by the
factors \(Q_{\rm dec}(ku)Q_{\rm dec}(kv)\) inside the radiation-era convolution.
For a narrow scalar peak and slowly varying \(Q_{\rm dec}\), this gives the
benchmark scaling
\(\Omega_{\rm GW}^{\rm eff}\simeq Q_{\rm dec}^2\Omega_{\rm GW}^{\rm class}\).
Thus quantum discord and decoherence provide a controlled way to characterize
the quantum-to-classical transition of PBH-producing perturbations, with the
clearest imprint appearing in scalar-induced gravitational waves.
\end{abstract}

\maketitle

\section{Introduction}
\label{sec:intro}
Primordial black holes (PBHs) offer a rare observational handle on the
small-scale primordial Universe \cite{Bird:2022wvk}. The cosmic microwave background (CMB)
probes curvature perturbations on very large scales, but PBHs are sensitive
to much shorter wavelengths that re-enter the Hubble radius during the
radiation era. If the primordial curvature perturbation is sufficiently
enhanced on these scales, rare overdense regions may collapse
gravitationally and form black holes. This makes PBHs a useful probe of
inflationary dynamics far beyond the CMB window
\cite{Carr:1974nx,Sasaki:2018dmp,Green:2020jor,Carr:2020gox,Carr:2021bzv}.
At the same time, PBH formation is not fixed by the power spectrum alone.
The collapse is a classical nonlinear process, depending on the density
contrast, the compaction function, the equation of state, and the shape of
the perturbation profile
\cite{Musco:2012au,Harada:2013epa,Escriva:2019phb,Musco:2020jjb}. This
distinction is important: the mechanism that generates large primordial
fluctuations belongs to inflationary physics, while the final collapse into
PBHs is governed by classical gravitational dynamics at horizon re-entry.

One simple and efficient way to enhance the scalar power spectrum is to
include a short non-attractor phase during inflation. A typical example is
ultra-slow-roll (USR) evolution, where the inflaton potential becomes
locally very flat and the inflaton velocity is rapidly damped by the
expansion of the Universe \cite{Kinney:2005vj,Namjoo:2012aa}. During this
stage the slow-roll parameter decreases approximately as
$\epsilon\propto a^{-6}$, and the would-be decaying mode of the curvature
perturbation grows outside the Hubble radius. As a result, the curvature
perturbation can be amplified by many orders of magnitude on a narrow range
of scales. This provides a natural route to PBH formation and also sources
a stochastic scalar-induced gravitational-wave (SIGW) background when the
enhanced scalar modes re-enter the horizon during radiation domination
\cite{Di:2017ndc,Byrnes:2018txb,Byrnes:2021jka,Kohri:2018awv,
Inomata:2019ivs} For related higher-dimensional scenarios involving primordial black holes and scalar-induced gravitational waves, see Refs.~\cite{Ahmed:2026ybo,Ahmed:2026pjd}.

The SIGW signal is especially important because it provides an indirect
probe of the same small-scale scalar fluctuations responsible for PBH
production. Even if PBHs constitute only a small fraction of the dark
matter, the accompanying induced gravitational waves may be observable by
pulsar timing arrays, space-based detectors, or ground-based interferometers,
depending on the peak scale of the scalar spectrum
\cite{Ananda:2006af,Baumann:2007zm,Saito:2008jc,Domenech:2021ztg,
Yuan:2021qgz}. In this sense, PBHs and SIGWs form a
closely related pair of observables: PBHs test the rare tail of the density
fluctuations, while SIGWs probe the two-point scalar covariance through the
second-order tensor source.

There is, however, a quantum question behind this classical PBH/SIGW
picture. The primordial curvature perturbations that later seed PBHs are
believed to originate from quantum vacuum fluctuations generated during
inflation and subsequently stretched to cosmological scales
\cite{Mukhanov:1981xt,Guth:1982ec,Hawking:1982cz}.
At the level of linear cosmological perturbation theory, each pair of
opposite Fourier modes, $(\bm{k},-\bm{k})$, evolves into a two-mode
squeezed Gaussian quantum state due to the parametric amplification by the
expanding background
\cite{Grishchuk:1990bj,Albrecht:1992kf,Polarski:1995jg,Kiefer:1998qe}.
After horizon exit, the squeezing parameter becomes very large, and the
perturbations behave in many practical respects like a classical stochastic
field, although the covariance matrix continues to encode information about
their quantum origin. This issue is closely related to the
quantum-to-classical transition of inflationary perturbations, which has
been extensively studied using squeezing, decoherence, and the emergence of
classical stochastic correlations
\cite{Polarski:1995jg,Kiefer:1998qe,Zurek:2003zz}.

Quantum-information theory gives a useful language for making this question
more precise. Entanglement is one possible measure of quantumness, but it is
not the only one. In mixed Gaussian states, entanglement can disappear while
weaker quantum correlations remain. Gaussian quantum discord is designed to
capture this broader class of quantum correlations
\cite{Ollivier:2001fdq,Giorda:2010wpy,Adesso:2011efy,
Weedbrook:2011wxo}. In cosmology, such
covariance-matrix diagnostics have been used to study the quantum nature of
inflationary perturbations and their behavior under decoherence
\cite{Martin:2015qta,Martin:2021znx,Martin:2022kph,Micheli2022tld,
Micheli2025yux}. These ideas suggest that PBH-producing modes should not be
viewed only as a classical random field from the beginning; rather, their
quantum origin and subsequent decoherence may leave useful information in
the Gaussian covariance structure.

The aim of this work is to apply this viewpoint to the PBH and SIGW problem
in a conservative way. We do not propose that quantum discord replaces the
PBH collapse threshold. PBH formation remains a classical threshold process:
a sufficiently large density perturbation must re-enter the Hubble radius
and collapse. Instead, we use Gaussian quantum discord as a diagnostic of
the residual quantum correlations carried by the PBH-producing curvature
modes. This allows us to ask a sharper question: after decoherence, how much
of the original quantum structure of the squeezed state can survive, and can
the same open-system dynamics modify the scalar covariance that enters PBH
and SIGW observables?

To describe this effect, we treat the scalar perturbations as an open
quantum system interacting with environmental degrees of freedom. These
degrees of freedom may represent other light fields, short-wavelength modes,
tensor perturbations, or unobserved sectors. After tracing over the
environment, the reduced state of a given $(\bm k,-\bm k)$ pair becomes
mixed. In such a mixed state, entanglement may be erased while Gaussian
discord remains nonzero. At the same time, the scalar two-point covariance
may or may not be suppressed, depending on the microscopic decoherence
channel. For example, a pure dephasing channel can destroy phase coherence
without changing the equal-time power spectrum, whereas a dissipative
attenuation channel can reduce the covariance itself
\cite{Zurek:2003zz,Burgess:2006jn,Burgess:2014eoa,
Nelson:2016kjm}.

We parametrize the possible covariance-level effect by writing
\begin{equation}
{\cal P}^{\rm eff}_{\zeta}(k)
=
Q_{\rm dec}(k)\,
{\cal P}^{\rm class}_{\zeta}(k),
\end{equation}
where $Q_{\rm dec}(k)$ is a model-dependent covariance-survival factor. It
should not be identified with Gaussian discord itself. Discord diagnoses
residual quantum correlations in the Gaussian state, while $Q_{\rm dec}$
measures how much of the scalar two-point function survives in a specified
open-system channel. This distinction is central to our analysis. If the
scalar covariance is suppressed, the PBH abundance changes through the
smoothed density variance entering the classical collapse probability. The
SIGW spectrum is affected more directly, because the induced tensor source
is quadratic in scalar perturbations. In the radiation-era convolution, the
effective spectrum contains the product
$Q_{\rm dec}(ku)Q_{\rm dec}(kv)$. For a narrow scalar peak and slowly
varying $Q_{\rm dec}$, this reduces to the useful benchmark relation
\begin{equation}
\Omega_{\rm GW}^{\rm eff}
\simeq
Q_{\rm dec}^{2}\,
\Omega_{\rm GW}^{\rm class}.
\end{equation}
This scaling is therefore a consequence of scalar covariance suppression,
not a direct production of gravitational waves by quantum discord.

In this paper we connect three ingredients that are often discussed
separately: the squeezed quantum state of inflationary perturbations, the
open-system description of decoherence, and the classical PBH/SIGW
phenomenology produced by an enhanced scalar power spectrum. Recent studies
have also explored related questions, such as quantum memory, induced tensor
coherence, and residual discord in secondary gravitational-wave backgrounds
\cite{Ahmed2026bkq,Ahmed2026hlz}. Here we focus specifically on the
PBH-producing scalar modes and on how a covariance-level decoherence factor
can enter the PBH abundance and SIGW spectrum without changing the
classical collapse condition.

The paper is organized as follows. In Sec.~\ref{sec:usr} we review the
background dynamics of ultra-slow-roll inflation and the super-Hubble growth
of the curvature perturbation. In Sec.~\ref{sec:gaussian} we describe each
$(\bm k,-\bm k)$ pair as a two-mode squeezed Gaussian state and introduce
the corresponding entanglement and Gaussian-discord diagnostics. In
Sec.~\ref{sec:decoherence} we present the open-system description of
decoherence and derive the Gaussian loss-channel form of the covariance
matrix. In Sec.~\ref{sec:timing} we discuss the timing of decoherence
relative to horizon exit, reheating, and radiation-era horizon re-entry. In
Sec.~\ref{sec:pbh} we review the PBH abundance calculation and explain how a
covariance-survival factor can modify the variance entering the classical
collapse probability. In Sec.~\ref{sec:sigw} we derive the corresponding
modification of the scalar-induced gravitational-wave spectrum and specify
the conditions under which the approximate $Q_{\rm dec}^{2}$ scaling is
valid. Numerical illustrations are presented in Sec.~\ref{sec:numerical},
and we summarize our conclusions in Sec.~\ref{sec:conclusions}.

\section{Ultra-slow-roll enhancement}
\label{sec:usr}

\subsection{Background dynamics}
We first review the background dynamics of ultra-slow-roll inflation. The
main purpose of this section is to show how a short non-attractor phase can
strongly amplify the curvature perturbation on small scales, providing the
necessary conditions for primordial-black-hole formation and the associated
scalar-induced gravitational-wave signal
\cite{Tsamis:2003px,Kinney:2005vj}.
We consider a canonical single-field inflationary model minimally coupled to
gravity \cite{Liddle:1994dx,Baumann:2009ds}. The action is

\begin{equation}
S =
\int d^4x \sqrt{-g}
\left[
\frac{M_{\rm Pl}^2}{2} R
-\frac{1}{2} g^{\mu\nu}\partial_\mu \phi \partial_\nu \phi
-V(\phi)
\right].
\label{eq:single_field_action}
\end{equation}
Here $M_{\rm Pl}$ is the reduced Planck mass, $\phi$ is the inflaton field,
and $V(\phi)$ is the inflationary potential.  For a spatially flat FLRW
background,
\begin{equation}
ds^2 = -dt^2 + a^2(t)d{\bf x}^2,
\label{eq:flrw_metric}
\end{equation}
the homogeneous background equations are
\begin{equation}
3M_{\rm Pl}^2H^2
=
\frac{1}{2}\dot{\phi}^2+V(\phi),
\label{eq:friedmann_1}
\end{equation}
\begin{equation}
-2M_{\rm Pl}^2\dot H
=
\dot{\phi}^2,
\label{eq:friedmann_2}
\end{equation}
and
\begin{equation}
\ddot{\phi}+3H\dot{\phi}+V_{,\phi}=0.
\label{eq:kg_equation}
\end{equation}
The first equation is the Friedmann equation and fixes the expansion rate in
terms of the energy density of the inflaton.  The second equation shows that
the time variation of $H$ is controlled by the kinetic energy of the field.
The third equation is the Klein--Gordon equation for the homogeneous inflaton.
The term $3H\dot{\phi}$ is the Hubble-friction term, while $V_{,\phi}$ is the
force generated by the potential.
It is useful to introduce the Hubble slow-roll parameters
\begin{equation}
\epsilon
=
-\frac{\dot H}{H^2}
=
\frac{\dot{\phi}^2}{2H^2M_{\rm Pl}^2},
\label{eq:epsilon_def}
\end{equation}
and
\begin{equation}
\eta_H
=
\frac{\dot{\epsilon}}{H\epsilon}
=
\frac{d\ln\epsilon}{dN},
\label{eq:etaH_def}
\end{equation}
where $N=\ln a$ is the number of $e$-folds.  In ordinary slow-roll inflation,
$\epsilon$ varies slowly and $|\eta_H|\ll1$.  Ultra-slow-roll inflation is
different.  It is a non-attractor regime in which the potential becomes
locally very flat,
\begin{equation}
V_{,\phi}\simeq0.
\label{eq:flat_potential_usr}
\end{equation}
In this limit the inflaton equation of motion reduces to
\begin{equation}
\ddot{\phi}+3H\dot{\phi}\simeq0.
\label{eq:usr_equation}
\end{equation}
This equation has a simple physical meaning.  Since the potential force is
negligible, the field velocity is damped almost entirely by Hubble friction.
For approximately constant $H$, the solution is
\begin{equation}
\dot{\phi}\propto a^{-3}.
\label{eq:phidot_usr}
\end{equation}
Therefore the kinetic energy decreases very rapidly,
\begin{equation}
\dot{\phi}^2\propto a^{-6}.
\label{eq:kinetic_usr}
\end{equation}
Using the definition of $\epsilon$, this gives
\begin{equation}
\epsilon
=
\frac{\dot{\phi}^2}{2H^2M_{\rm Pl}^2}
\propto a^{-6}.
\label{eq:epsilon_usr}
\end{equation}
Since $N=\ln a$, we then obtain
\begin{equation}
\eta_H
=
\frac{d\ln\epsilon}{dN}
\simeq -6.
\label{eq:etaH_usr}
\end{equation}
Thus canonical ultra-slow-roll inflation is characterized by
\begin{equation}
\dot{\phi}\propto a^{-3},
\qquad
\epsilon\propto a^{-6},
\qquad
\eta_H\simeq -6.
\label{eq:usr_summary}
\end{equation}

This result is important because USR is not an attractor phase.  In ordinary
slow-roll inflation, the curvature perturbation becomes conserved after
horizon exit.  In USR, however, the would-be decaying mode grows on
super-Hubble scales.  This growth can strongly enhance the curvature power
spectrum on small scales, making USR a natural mechanism for producing PBHs
and scalar-induced gravitational waves
\cite{Kinney:2005vj,Namjoo:2012aa,Byrnes:2018txb,Byrnes:2021jka}.

Finally, let us clarify the convention used for the second slow-roll
parameter.  In this work we define
\begin{equation}
\eta_H
=
\frac{d\ln\epsilon}{dN}.
\label{eq:etaH_convention}
\end{equation}
With this definition, canonical USR gives $\eta_H\simeq -6$.  Some papers
instead use the field-acceleration parameter
\begin{equation}
\eta_\phi
=
-\frac{\ddot{\phi}}{H\dot{\phi}},
\label{eq:eta_phi_def}
\end{equation}
for which USR gives $\eta_\phi\simeq3$.  These two statements are not in
conflict; they simply correspond to different definitions.

\subsection{Growth of the curvature perturbation}
  For a canonical single-field model, the scalar perturbations
are conveniently described in terms of the Mukhanov--Sasaki variable
$v_k=z\zeta_k$, where
\begin{equation}
z=a\sqrt{2\epsilon}\,M_{\rm Pl}.
\label{eq:z_def}
\end{equation}
Here $\zeta_k$ is the comoving curvature perturbation and a prime denotes a
derivative with respect to conformal time $\tau$.  The variable $v_k$ obeys
the Mukhanov--Sasaki equation
\begin{equation}
v_k''
+
\left(
k^2-\frac{z''}{z}
\right)v_k
=
0.
\label{eq:ms_equation}
\end{equation}
Equivalently, using $v_k=z\zeta_k$, one obtains the equation for the curvature
perturbation,
\begin{equation}
\zeta_k''
+
2\frac{z'}{z}\zeta_k'
+
k^2\zeta_k
=
0.
\label{eq:zeta_equation}
\end{equation}
These equations are standard in the theory of inflationary perturbations
\cite{Mukhanov:1985rz,Sasaki:1986hm,Mukhanov:1990me}.

On super-Hubble scales, $k\ll aH$, the gradient term can be neglected.  In
cosmic time this gives
\begin{equation}
\frac{d}{dt}
\left(
a^3\epsilon\dot{\zeta}_k
\right)
\simeq
0.
\label{eq:zeta_conservation_form}
\end{equation}
Therefore the general super-Hubble solution can be written as
\begin{equation}
\zeta_k(t)
=
C_1
+
C_2
\int^t
\frac{dt'}{a^3(t')\epsilon(t')}.
\label{eq:zeta_super_solution}
\end{equation}
The first term is the constant mode, while the second term is the mode that
usually decays in ordinary slow-roll inflation.

In ordinary slow roll, $\epsilon$ is approximately constant.  Hence
\begin{equation}
\int^t
\frac{dt'}{a^3(t')\epsilon(t')}
\propto
a^{-3},
\label{eq:sr_decaying_mode}
\end{equation}
so the second mode decays and $\zeta_k$ becomes conserved after horizon exit.
This is the usual attractor behavior of single-field slow-roll inflation.

In USR, the situation is different.  Since $\epsilon\propto a^{-6}$, the
integrand in Eq.~\eqref{eq:zeta_super_solution} behaves as
\begin{equation}
\frac{1}{a^3\epsilon}
\propto
a^3.
\label{eq:usr_integrand}
\end{equation}
For approximately constant $H$, this gives
\begin{equation}
\int^t
\frac{dt'}{a^3(t')\epsilon(t')}
\propto
\int^t dt'\,a^3(t')
\propto
a^3.
\label{eq:usr_growing_integral}
\end{equation}
Thus the would-be decaying mode becomes a growing mode during USR,
\begin{equation}
\zeta_k
\propto
a^3.
\label{eq:zeta_growth_usr}
\end{equation}
If the USR phase lasts for $\Delta N$ $e$-folds, the curvature perturbation is
therefore enhanced as
\begin{equation}
\zeta_k
\propto
e^{3\Delta N}.
\label{eq:zeta_growth_efolds}
\end{equation}
Since the power spectrum is quadratic in $\zeta_k$, the corresponding
enhancement of the curvature power spectrum is
\begin{equation}
{\cal P}_{\zeta}(k)
\propto
e^{6\Delta N}.
\label{eq:power_growth_usr}
\end{equation}

This super-Hubble growth is the basic mechanism by which a short USR stage can
generate a large peak in the small-scale curvature power spectrum.  Such a
peak can seed PBH formation after horizon re-entry and also source a
scalar-induced gravitational-wave background at second order
\cite{Kinney:2005vj,Namjoo:2012aa,Byrnes:2018txb,Byrnes:2021jka}.  In a
realistic model, the exact height and shape of the peak also depend on the
duration of the USR phase, the transition into and out of USR, and the
matching of the perturbation modes across these phases.  Therefore
Eq.~\eqref{eq:power_growth_usr} should be understood as the idealized scaling
for an approximately constant-$H$ USR stage, not as a complete numerical
prediction for every mode.

\section{Gaussian quantum state and discord}
\label{sec:gaussian}

\subsection{Two-mode squeezed vacuum}

We now describe the quantum state of the scalar perturbations.  At linear
order, each Fourier mode of the curvature perturbation behaves as a quantum
harmonic oscillator with a time-dependent frequency.  Because of spatial
translation invariance, modes with opposite momenta, $\bm k$ and $-\bm k$,
are produced in correlated pairs.  Starting from the Bunch--Davies vacuum,
the late-time state of each pair is well described by a two-mode squeezed
Gaussian state \cite{Polarski:1995jg,Kiefer:1998qe,Martin:2015qta}.

For each pair $(\bm k,-\bm k)$, we write the state as
\begin{equation}
|\psi_{\bm k}\rangle
=
\frac{1}{\cosh r_k}
\sum_{n=0}^{\infty}
\left(
-e^{i\varphi_k}\tanh r_k
\right)^n
|n_{\bm k},n_{-\bm k}\rangle .
\label{eq:tmsv_state}
\end{equation}
Here $r_k$ is the squeezing parameter and $\varphi_k$ is the squeezing phase.
This is the standard two-mode squeezed form of inflationary perturbations,
where opposite Fourier modes are correlated by the time-dependent background
\cite{Grishchuk:1990bj,Albrecht:1992kf,Polarski:1995jg,Kiefer:1998qe,
Martin:2015qta}. The parameter $r_k$ measures how strongly the two modes are
correlated. The mean occupation number of each mode is
\begin{equation}
\bar n_k
=
\sinh^2 r_k .
\label{eq:occupation_number}
\end{equation}
Thus, when $r_k$ becomes large, the state contains many correlated quanta.
This is one reason why inflationary perturbations can look classical after
horizon exit, although their origin is quantum mechanical.

To describe the state in phase space, we introduce the quadrature operators
\begin{equation}
\hat q_j
=
\frac{\hat a_j+\hat a_j^\dagger}{\sqrt{2}},
\label{eq:q_quadrature}
\end{equation}
and
\begin{equation}
\hat p_j
=
\frac{\hat a_j-\hat a_j^\dagger}{i\sqrt{2}}.
\label{eq:p_quadrature}
\end{equation}
They obey the canonical commutation relation
\begin{equation}
[\hat q_i,\hat p_j]
=
i\delta_{ij}.
\label{eq:canonical_commutator}
\end{equation}
Using the quadrature vector
\begin{equation}
\hat{\bm R}
=
(\hat q_{\bm k},\hat p_{\bm k},
 \hat q_{-\bm k},\hat p_{-\bm k})^T ,
\label{eq:quadrature_vector}
\end{equation}
the covariance matrix is defined by
\begin{equation}
V_{ij}
=
\frac{1}{2}
\left\langle
\hat R_i\hat R_j+\hat R_j\hat R_i
\right\rangle
-
\left\langle \hat R_i\right\rangle
\left\langle \hat R_j\right\rangle .
\label{eq:covariance_definition}
\end{equation}
These phase-space conventions are standard in continuous-variable Gaussian
quantum information \cite{Weedbrook:2011wxo}.
For a convenient choice of squeezing phase, the covariance matrix of the
two-mode squeezed vacuum takes the standard form
\begin{equation}
V_{\rm TMSV}
=
\frac{1}{2}
\begin{pmatrix}
\cosh 2r & 0 & \sinh 2r & 0 \\
0 & \cosh 2r & 0 & -\sinh 2r \\
\sinh 2r & 0 & \cosh 2r & 0 \\
0 & -\sinh 2r & 0 & \cosh 2r
\end{pmatrix}.
\label{eq:tmsv_covariance}
\end{equation}
For a general squeezing phase $\varphi_k$, this matrix is rotated in phase
space. However, the symplectic eigenvalues and entropy-based correlation
measures are unchanged by such a local phase-space rotation
\cite{Weedbrook:2011wxo}.
With the convention used above, the vacuum covariance matrix is
\begin{equation}
V_{\rm vac}
=
\frac{1}{2}I_4 .
\label{eq:vacuum_covariance}
\end{equation}
Therefore the vacuum symplectic eigenvalue is $1/2$. The full two-mode
squeezed state in Eq.~\eqref{eq:tmsv_state} is pure, and its two symplectic
eigenvalues are
\begin{equation}
\nu_+
=
\nu_-
=
\frac{1}{2}.
\label{eq:pure_symplectic_eigenvalues}
\end{equation}
However, if one traces over the partner mode $-\bm k$, the remaining single
mode is mixed. Its reduced covariance matrix is
\begin{equation}
V_{\bm k}
=
\frac{1}{2}\cosh 2r_k\, I_2 .
\label{eq:reduced_covariance}
\end{equation}
This reduced mixedness is the origin of the entanglement entropy between the
two modes, and for a pure bipartite Gaussian state it is also equal to the
Gaussian quantum discord
\cite{Giorda:2010wpy,Adesso:2011efy,Weedbrook:2011wxo,Martin:2015qta}.
\subsection{Pure-state discord}

We next compute the quantum discord of the pure two-mode squeezed state.  For
a single bosonic mode with occupation number $n$, the von Neumann entropy is
\begin{equation}
g(n)
=
(n+1)\log_2(n+1)
-
n\log_2 n .
\label{eq:boson_entropy}
\end{equation}
The logarithm is taken in base two, so the entropy is measured in bits.

For a pure bipartite state, the quantum discord is equal to the entropy of
either reduced subsystem.  Therefore, for the pure two-mode squeezed state, we
obtain
\begin{equation}
\mathcal D_G^{\rm pure}(r)
=
g(\sinh^2 r).
\label{eq:pure_discord}
\end{equation}
This result is simple but important.  In a pure squeezed state, large
squeezing automatically implies large quantum discord.  Thus pure-state
discord mainly tracks the amount of squeezing; it does not by itself provide
a new PBH collapse criterion.

For illustration, one finds
\begin{equation}
\mathcal D_G(0.2)
\simeq
0.247,
\label{eq:discord_example_02}
\end{equation}
\begin{equation}
\mathcal D_G(1.0)
\simeq
2.337,
\label{eq:discord_example_10}
\end{equation}
and
\begin{equation}
\mathcal D_G(1.3)
\simeq
3.196.
\label{eq:discord_example_13}
\end{equation}
A low threshold such as $\mathcal D_G^{\rm th}=0.2$ bits is already crossed
at approximately
\begin{equation}
r_{\rm th}
\simeq
0.175 .
\label{eq:discord_threshold_r}
\end{equation}
This shows that, in the pure-state limit, PBH-producing modes are expected to
have large discord once they are strongly squeezed.  Therefore the role of
discord in this work is not to replace the classical PBH threshold.  Instead,
discord is used as a diagnostic of residual quantum correlations, especially
after decoherence has made the state mixed \cite{Giorda:2010wpy,Adesso:2011efy,
Martin:2015qta}.
\section{Microscopic model of decoherence}
\label{sec:decoherence}

\subsection{System-environment interaction}

The pure two-mode squeezed state discussed above is an idealized description
of the scalar perturbations in a closed system.  The physically nontrivial
case arises when the PBH-producing curvature modes are treated as an open
quantum system.  In that case, the relevant pair $(\bm k,-\bm k)$ is not
described by a pure state, but by a reduced density matrix obtained after
tracing over environmental degrees of freedom.  These environmental modes may
represent other light fields, short-wavelength scalar modes, tensor modes, or
unobserved sub-Hubble fluctuations.  Such open-system descriptions have been
used extensively to study the quantum-to-classical transition of inflationary
perturbations and the possible observational consequences of decoherence
\cite{Burgess:2006jn,Martineau:2006ki,Nelson:2016kjm,
Martin:2018zbe,Burgess:2022nwu}.

We model the environment as a collection of harmonic oscillators,
\begin{equation}
H_{\rm env}
=
\sum_{\lambda}
\omega_\lambda
b_\lambda^\dagger b_\lambda ,
\label{eq:env_hamiltonian}
\end{equation}
and assume that the curvature-mode pair interacts weakly with this bath.  The
total Hamiltonian can be written schematically as
\begin{equation}
H_{\rm tot}
=
H_{\rm sys}
+
H_{\rm env}
+
H_{\rm int},
\label{eq:total_hamiltonian}
\end{equation}
where $H_{\rm sys}$ describes the two-mode scalar system and $H_{\rm int}$
contains the system-environment coupling.  For example, a linear coupling may
be written schematically as
\begin{equation}
H_{\rm int}
=
\sum_{\lambda}
g_\lambda
\hat X
\left(
b_\lambda+b_\lambda^\dagger
\right),
\label{eq:interaction_hamiltonian}
\end{equation}
where $\hat X$ is a linear combination of the system quadratures and
$g_\lambda$ are coupling constants.  More complicated gravitational or
self-interaction channels can also lead to decoherence, but the simple linear
model is sufficient to motivate the Gaussian channel used below.

After tracing over the environment, the reduced density matrix of the two-mode
system is
\begin{equation}
\rho_{\rm sys}(t)
=
{\rm Tr}_{\rm env}\,
\rho_{\rm tot}(t).
\label{eq:reduced_density_matrix}
\end{equation}
In the weak-coupling and Markovian limit, the reduced dynamics can be written
in Lindblad form,
\begin{equation}
\frac{d\rho_{\rm sys}}{dt}
=
-i
\left[
H_{\rm eff},
\rho_{\rm sys}
\right]
+
\sum_j
\gamma_j
\left(
L_j\rho_{\rm sys}L_j^\dagger
-
\frac{1}{2}
\left\{
L_j^\dagger L_j,
\rho_{\rm sys}
\right\}
\right).
\label{eq:lindblad_master}
\end{equation}
Here $H_{\rm eff}$ is the effective system Hamiltonian, $L_j$ are Lindblad
jump operators, and $\gamma_j$ are positive rates determined by the coupling
strengths and the bath spectral density.  This equation preserves positivity
and trace of the density matrix, and it gives a controlled phenomenological
description of dissipative and decohering dynamics
\cite{Breuer:2002pc,Breuer:2003avm}.

For a Gaussian system, it is convenient to work with the quadrature vector
\begin{equation}
\hat{\bm R}
=
(\hat q_{\bm k},\hat p_{\bm k},
 \hat q_{-\bm k},\hat p_{-\bm k})^T .
\label{eq:quadrature_vector_decoh}
\end{equation}
If the Hamiltonian is quadratic in the quadratures and the Lindblad operators
are linear in the quadratures, then Gaussianity is preserved.  In that case,
the complete state is characterized by its first moments and covariance
matrix.  Since the first moments vanish for the states considered here, the
relevant information is contained in the covariance matrix alone.

\subsection{Covariance matrix evolution}

The covariance matrix is defined as
\begin{equation}
V_{ij}
=
\frac{1}{2}
\left\langle
\hat R_i\hat R_j+\hat R_j\hat R_i
\right\rangle
-
\left\langle \hat R_i\right\rangle
\left\langle \hat R_j\right\rangle .
\label{eq:covariance_matrix_decoh}
\end{equation}
For a Gaussian Lindblad equation, $V$ obeys a linear Lyapunov equation,
\begin{equation}
\frac{dV}{dt}
=
K V
+
V K^T
+
D .
\label{eq:lyapunov}
\end{equation}
Here $K$ is the drift matrix and $D$ is the diffusion matrix.  Both are fixed
by the effective Hamiltonian and the Lindblad operators.  The drift matrix
describes the deterministic damping and rotation of the phase-space
variables, while the diffusion matrix describes noise injected by the
environment.

We write the covariance matrix in block form as
\begin{equation}
V
=
\begin{pmatrix}
V_A & V_C \\
V_C^T & V_B
\end{pmatrix},
\label{eq:block_covariance}
\end{equation}
where $V_A$ and $V_B$ are the local covariance matrices of the two modes
$\bm k$ and $-\bm k$, while $V_C$ contains their correlations.  For a local
attenuation channel acting identically on the two modes, the solution of
Eq.~\eqref{eq:lyapunov} can be written as
\begin{equation}
V_A^{(\eta)}
=
\eta V_A^{(0)}
+
(1-\eta)
\left(
n_{\rm env}
+
\frac{1}{2}
\right)
I_2 ,
\label{eq:loss_channel_A}
\end{equation}
\begin{equation}
V_B^{(\eta)}
=
\eta V_B^{(0)}
+
(1-\eta)
\left(
n_{\rm env}
+
\frac{1}{2}
\right)
I_2 ,
\label{eq:loss_channel_B}
\end{equation}
and
\begin{equation}
V_C^{(\eta)}
=
\eta V_C^{(0)} .
\label{eq:loss_channel_C}
\end{equation}
Here $V_A^{(0)}$, $V_B^{(0)}$, and $V_C^{(0)}$ are the initial blocks of the
pure two-mode squeezed covariance matrix in Eq.~\eqref{eq:block_covariance}.
The parameter $\eta$ is the transmissivity of the Gaussian attenuator channel.
For a constant damping rate $\Gamma$, one has
\begin{equation}
\eta(t)
=
e^{-\Gamma t}.
\label{eq:eta_definition}
\end{equation}
The environmental occupation number is
\begin{equation}
n_{\rm env}
=
\frac{1}{\exp(\omega/T_{\rm env})-1},
\label{eq:n_env}
\end{equation}
where $\omega\simeq k/a$ is the physical frequency of the mode and
$T_{\rm env}$ is the effective temperature of the environment.

Equations~\eqref{eq:loss_channel_A}--\eqref{eq:loss_channel_C} should be
understood as a specific dissipative Gaussian channel, not as the most general
form of cosmological decoherence.  This distinction is important.  Pure
dephasing can suppress off-diagonal elements of the density matrix without
changing the equal-time power spectrum, whereas an attenuation channel can
also reduce the covariance.  Therefore the covariance-suppression factor used
later in the PBH and SIGW calculations is a model-dependent quantity.  In the
present setup, the effective scalar power spectrum may be parametrized as
\begin{equation}
{\cal P}_{\zeta}^{\rm eff}(k)
=
Q_{\rm dec}(k)
{\cal P}_{\zeta}^{\rm class}(k),
\label{eq:P_eff_Qdec}
\end{equation}
where $Q_{\rm dec}(k)$ measures the survival of the scalar two-point
covariance in the chosen open-system channel.  In a cold-environment
attenuator channel, $Q_{\rm dec}(k)$ is approximately controlled by
$\eta(k)$, while in a pure dephasing channel one can have $Q_{\rm dec}=1$ for
the power spectrum even though the state has decohered.

\subsection{Entanglement and discord}

We now describe how entanglement and Gaussian discord are computed from the
mixed covariance matrix.  For a two-mode Gaussian state written in the block
form of Eq.~\eqref{eq:block_covariance}, the symplectic eigenvalues are
\begin{equation}
\nu_\pm
=
\sqrt{
\frac{
\Delta
\pm
\sqrt{
\Delta^2
-
4\det V
}
}{2}
},
\label{eq:symplectic_eigenvalues}
\end{equation}
where
\begin{equation}
\Delta
=
\det V_A
+
\det V_B
+
2\det V_C .
\label{eq:Delta_symplectic}
\end{equation}
Physical covariance matrices satisfy $\nu_\pm\geq 1/2$.

Entanglement is diagnosed by the partially transposed covariance matrix.  At
the covariance-matrix level, partial transposition corresponds to reversing
the sign of one momentum quadrature.  The corresponding symplectic invariant
is
\begin{equation}
\widetilde{\Delta}
=
\det V_A
+
\det V_B
-
2\det V_C .
\label{eq:Delta_partial_transpose}
\end{equation}
The smallest partially transposed symplectic eigenvalue is therefore
\begin{equation}
\widetilde{\nu}_-
=
\sqrt{
\frac{
\widetilde{\Delta}
-
\sqrt{
\widetilde{\Delta}^2
-
4\det V
}
}{2}
}.
\label{eq:nu_tilde_minus}
\end{equation}
The state is entangled when
\begin{equation}
\widetilde{\nu}_-
<
\frac{1}{2}.
\label{eq:ppt_entanglement_condition}
\end{equation}
The logarithmic negativity is then
\begin{equation}
E_N
=
\max
\left[
0,
-\log_2
\left(
2\widetilde{\nu}_-
\right)
\right].
\label{eq:log_negativity}
\end{equation}
This quantity measures entanglement, and it vanishes once the state becomes
separable \cite{Vidal:2002zz,Plenio:2005cwa,Adesso:2005}.

Gaussian discord is more general than entanglement.  It can remain nonzero
even after the logarithmic negativity has vanished.  To compute it, one
minimizes the conditional entropy over Gaussian measurements on one of the two
modes.  For the symmetric states used in our numerical examples, we may take
the measurement covariance matrix as
\begin{equation}
\Gamma(s)
=
\frac{1}{2}
\begin{pmatrix}
s & 0 \\
0 & 1/s
\end{pmatrix},
\qquad
s>0 .
\label{eq:measurement_covariance}
\end{equation}
The conditional covariance matrix of mode $A$ after a Gaussian measurement on
mode $B$ is
\begin{equation}
V_{A|B}(s)
=
V_A
-
V_C
\left[
V_B+\Gamma(s)
\right]^{-1}
V_C^T .
\label{eq:conditional_covariance}
\end{equation}

For a one-mode Gaussian state with symplectic eigenvalue $\nu$, the entropy is
\begin{equation}
h(\nu)
=
\left(
\nu+\frac{1}{2}
\right)
\log_2
\left(
\nu+\frac{1}{2}
\right)
-
\left(
\nu-\frac{1}{2}
\right)
\log_2
\left(
\nu-\frac{1}{2}
\right).
\label{eq:h_entropy_nu}
\end{equation}
The total entropy of the two-mode state is
\begin{equation}
S(AB)
=
h(\nu_+)+h(\nu_-).
\label{eq:entropy_AB}
\end{equation}
Similarly, the entropy of mode $B$ is
\begin{equation}
S(B)
=
h\left(\sqrt{\det V_B}\right).
\label{eq:entropy_B}
\end{equation}
The conditional entropy after measurement is
\begin{equation}
S(A|B_s)
=
h
\left(
\sqrt{
\det V_{A|B}(s)
}
\right).
\label{eq:conditional_entropy}
\end{equation}
The Gaussian discord with measurement on $B$ is then
\begin{equation}
{\cal D}_G^{A|B}
=
S(B)
-
S(AB)
+
\min_{s>0}
S(A|B_s).
\label{eq:gaussian_discord_mixed}
\end{equation}
In the pure limit, $\nu_+=\nu_-=1/2$ and this expression reduces to
Eq.~\eqref{eq:pure_discord}.  Therefore, the mixed-state calculation smoothly
connects to the pure two-mode squeezed result discussed in the previous
section.

This formalism makes the physical interpretation transparent.  The
logarithmic negativity $E_N$ tells us whether the two curvature modes are
still entangled.  Gaussian discord ${\cal D}_G$ instead measures quantum
correlations more broadly, and can remain nonzero even when the state is
already separable.  This is why the decohered mixed state is the genuinely
interesting regime for the present work \cite{Giorda:2010wpy,Adesso:2011efy,
Weedbrook:2011wxo}.

\section{Timing of decoherence and its observational consequences}
\label{sec:timing}

The effect of decoherence depends not only on its strength, but also on when it
acts.  For PBH formation and scalar-induced gravitational waves, the relevant
times are horizon exit during inflation, the possible USR phase, reheating, and
horizon re-entry during radiation domination.  These stages are physically
different, and a single time-independent decoherence parameter cannot capture
all possibilities.

In ordinary slow-roll inflation, the curvature perturbation $\zeta_k$ becomes
conserved on super-Hubble scales.  In USR, however, the background is
non-attractor and $\zeta_k$ can grow outside the Hubble radius.  Therefore the
quantum-to-classical transition of the PBH-producing modes should be discussed
together with the background evolution.  Decoherence can begin around horizon
exit, but in principle it may continue during the super-Hubble evolution,
reheating, and even after the mode has re-entered the Hubble radius.  The
interaction with environmental degrees of freedom is not switched off sharply
at horizon crossing \cite{Polarski:1995jg,Kiefer:1998qe,Burgess:2006jn,
Nelson:2016kjm,Martin:2018zbe}.

We parametrize the cumulative effect of the environment by an effective
transmissivity $\eta_{\rm eff}(k)$ and an effective environmental occupation
$n_{\rm env}^{\rm eff}(k)$.  These quantities summarize the open-system
evolution from an initial time $\tau_i$ to a final time $\tau_f$.  For example,
if the damping rate is time-dependent, one may write
\begin{equation}
\eta_{\rm eff}(k)
=
\exp
\left[
-
\int_{t_i}^{t_f}
\Gamma(k,t)\,dt
\right].
\label{eq:eta_eff_timing}
\end{equation}
Here $\Gamma(k,t)$ is the effective decoherence or damping rate.  In a
cosmological setting, $\Gamma(k,t)$ can depend on the Hubble scale $H(t)$, the
physical momentum $k/a(t)$, the temperature of the environment, and the
strength of the system-environment coupling.

It is useful to distinguish the physical momentum from the horizon-crossing
condition.  The physical momentum is
\begin{equation}
p_{\rm phys}(t)
=
\frac{k}{a(t)} ,
\label{eq:physical_momentum}
\end{equation}
and it decreases as the universe expands.  However, during radiation
domination the quantity $aH$ decreases, so the ratio
\begin{equation}
\frac{k}{aH}
\label{eq:horizon_ratio}
\end{equation}
increases with time.  This is why modes that were outside the Hubble radius
eventually re-enter.  Larger $k$ modes, or equivalently higher-frequency modes,
re-enter earlier during radiation domination.

For PBH formation, the relevant final time is approximately the horizon
re-entry time of the scale $k$, defined by
\begin{equation}
k
=
a_f H_f .
\label{eq:horizon_reentry_condition}
\end{equation}
At this time, the classical collapse criterion is applied to the density
contrast or compaction function.  Decoherence does not replace this collapse
criterion.  It can affect the PBH abundance only if the open-system dynamics
also modifies the scalar two-point covariance.  In that case, we may write
\begin{equation}
{\cal P}_{\zeta}^{\rm eff}(k)
=
Q_{\rm dec}(k)
{\cal P}_{\zeta}^{\rm class}(k),
\label{eq:Pzeta_eff_timing}
\end{equation}
where $Q_{\rm dec}(k)$ is a model-dependent covariance-survival factor.  In a
pure dephasing channel, one may have $Q_{\rm dec}=1$ even though the reduced
density matrix has decohered.  In a dissipative attenuation channel,
$Q_{\rm dec}<1$ is possible.

The PBH abundance is exponentially sensitive to the variance of the smoothed
perturbation.  Schematically, for a Gaussian estimate one has
\begin{equation}
\beta_f(M)
\simeq
\frac{1}{2}
{\rm erfc}
\left(
\frac{\delta_c}{\sqrt{2}\sigma_\delta}
\right),
\label{eq:beta_timing}
\end{equation}
where $\delta_c$ is the collapse threshold and $\sigma_\delta$ is the smoothed
density variance.  Therefore, if the covariance is reduced by
$Q_{\rm dec}(k)$, the PBH abundance can be strongly suppressed.  Even a modest
change in the variance can produce orders-of-magnitude changes in
$f_{\rm PBH}$.

For scalar-induced gravitational waves, the timing issue is slightly more
subtle.  The tensor perturbations are sourced continuously during the
radiation era by products of scalar perturbations.  Therefore, in the most
general case, the decoherence factor should depend on both scale and time,
\begin{equation}
Q_{\rm dec}
=
Q_{\rm dec}(k,\tau).
\label{eq:Qdec_time_dependent}
\end{equation}
The induced gravitational-wave spectrum would then have to be computed using
the time-dependent scalar covariance inside the radiation-era source integral.
This would require solving the open-system evolution and the second-order
tensor equation simultaneously.

In the present work we restrict ourselves to a simpler saturation regime.  We
assume that the relevant decoherence process is completed before, or soon
after, the scalar modes begin to source the tensor perturbations efficiently.
In this limit, the time-dependent factor can be approximated by a static
quantity,
\begin{equation}
Q_{\rm dec}(k,\tau)
\simeq
Q_{\rm dec}(k).
\label{eq:Qdec_static_approx}
\end{equation}
This approximation is appropriate when the decoherence timescale is shorter
than the Hubble timescale around horizon re-entry.  Under this assumption, the
main effect of decoherence is to rescale the scalar power spectrum entering
the PBH abundance and the scalar-induced gravitational-wave convolution.

Thus the observational consequences can be summarized as follows.  If the
environment only decoheres phases but leaves the scalar covariance unchanged,
then PBH abundance and SIGWs are essentially unaffected at the level of the
power spectrum.  If, however, the environment acts as a dissipative Gaussian
channel and reduces the scalar covariance, then PBH production is suppressed
through the variance, while the SIGW signal is suppressed through the product
of two scalar power spectra.  For a narrow scalar peak and slowly varying
$Q_{\rm dec}(k)$, this gives the approximate scaling
\begin{equation}
\Omega_{\rm GW}^{\rm eff}
\simeq
Q_{\rm dec}^2
\Omega_{\rm GW}^{\rm class}.
\label{eq:Qdec2_timing}
\end{equation}
This is the regime explored in the numerical examples below.

\section{PBH abundance without and with quantum-information diagnostics}
\label{sec:pbh}

\subsection{Classical PBH abundance}

PBH formation is a classical gravitational-collapse process. A perturbation
collapses into a PBH when its density contrast, or equivalently its compaction
function, exceeds a threshold value at horizon re-entry. The threshold is not
universal: it depends on the shape of the perturbation profile and on the
equation of state of the background fluid. For radiation domination, typical
values are $\delta_c\simeq 0.4$--$0.6$
\cite{Carr:1974nx,Musco:2012au,Harada:2013epa,Escriva:2019phb,Musco:2020jjb}.

For a Gaussian estimate, one smooths the radiation-era density contrast on a
comoving scale $R$. The variance is
\begin{equation}
\sigma_\delta^2(R)
=
\int d\ln k\,
W^2(kR)
\left[
\frac{4}{9}
(kR)^2
T(kR)
\right]^2
{\cal P}_{\zeta}(k).
\label{eq:sigma_delta}
\end{equation}
Here $W(kR)$ is a window function and $T(kR)$ is the radiation-era transfer
function. The factor $(4/9)(kR)^2$ relates the density contrast to the
curvature perturbation during radiation domination. This form is commonly
used in PBH abundance estimates based on the smoothed density contrast
\cite{Sasaki:2018dmp,Green:2020jor,Carr:2020gox}.

Assuming a Gaussian probability distribution for the smoothed density contrast,
\begin{equation}
P(\delta_R)
=
\frac{1}{\sqrt{2\pi}\sigma_\delta}
\exp
\left[
-\frac{\delta_R^2}{2\sigma_\delta^2}
\right],
\label{eq:gaussian_pdf_delta}
\end{equation}
the initial PBH formation fraction is
\begin{equation}
\beta_f(M)
=
\int_{\delta_c}^{\infty}
P(\delta_R)\,d\delta_R .
\label{eq:beta_integral}
\end{equation}
This gives
\begin{equation}
\beta_f(M)
=
\frac{1}{2}
{\rm erfc}
\left(
\frac{\delta_c}{\sqrt{2}\sigma_\delta}
\right).
\label{eq:beta}
\end{equation}
For $\sigma_\delta\ll\delta_c$, the abundance is exponentially sensitive to
the variance,
\begin{equation}
\beta_f(M)
\simeq
\frac{\sigma_\delta}{\sqrt{2\pi}\delta_c}
\exp
\left[
-\frac{\delta_c^2}{2\sigma_\delta^2}
\right].
\label{eq:beta_asymptotic}
\end{equation}
This exponential dependence is the reason why even a small change in the
small-scale power spectrum can produce a very large change in the PBH
abundance \cite{Carr:1974nx,Sasaki:2018dmp,Green:2020jor}.

The present PBH dark-matter fraction is approximately
\begin{equation}
f_{\rm PBH}(M)
\simeq
6.3\times 10^{15}
\left(
\frac{\gamma}{0.2}
\right)^{1/2}
\left(
\frac{g_*}{106.75}
\right)^{-1/4}
\left(
\frac{M}{10^{18}{\rm g}}
\right)^{-1/2}
\beta_f(M),
\label{eq:fpbh}
\end{equation}
where $\gamma$ is the collapse efficiency factor and $g_*$ is the number of
relativistic degrees of freedom at formation. Equivalently, one may write
\begin{equation}
\beta_f(M)
\simeq
1.6\times 10^{-16}
\left(
\frac{\gamma}{0.2}
\right)^{-1/2}
\left(
\frac{g_*}{106.75}
\right)^{1/4}
\left(
\frac{M}{10^{18}{\rm g}}
\right)^{1/2}
f_{\rm PBH}(M).
\label{eq:beta_from_fpbh}
\end{equation}
These relations are the standard conversion between the initial PBH mass
fraction at formation and the present PBH dark-matter fraction
\cite{Sasaki:2018dmp,Green:2020jor,Carr:2020gox,Carr:2021bzv}.

The PBH mass is related to the comoving scale that re-enters the Hubble radius
during radiation domination. For the conventions used here we take
\begin{equation}
M(k)
\simeq
10^{18}\,{\rm g}\,
\left(\frac{\gamma}{0.2}\right)
\left(\frac{g_*}{106.75}\right)^{-1/6}
\left(
\frac{k}{7.0\times10^{13}\,{\rm Mpc}^{-1}}
\right)^{-2}.
\label{eq:mass_wavenumber_relation}
\end{equation}
Equivalently,
\begin{equation}
k_M
\simeq
7.0\times10^{13}\,{\rm Mpc}^{-1}
\left(\frac{\gamma}{0.2}\right)^{1/2}
\left(\frac{g_*}{106.75}\right)^{-1/12}
\left(
\frac{M}{10^{18}\,{\rm g}}
\right)^{-1/2}.
\label{eq:wavenumber_mass_relation}
\end{equation}
Thus larger comoving wavenumbers correspond to earlier horizon re-entry and
smaller PBH masses. This mass--scale map is standard in PBH phenomenology and
is used below when displaying the PBH abundance as a function of mass
\cite{Sasaki:2018dmp,Kohri:2018awv,Inomata:2019ivs,Domenech:2021ztg}.

\subsection{Quantum-information diagnostics}

We now explain how the quantum-information sector can be included without
changing the classical nature of PBH collapse.  The key point is that quantum
discord does not replace the threshold $\delta_c$.  PBH formation is still
controlled by the condition that the density perturbation exceeds the
collapse threshold at horizon re-entry.  Gaussian discord instead provides a
diagnostic of the quantum correlations carried by the curvature modes that
seed the PBH.

If the covariance-matrix parameters are deterministic functions of the scale
$k$, then a simple way to impose a diagnostic condition is
\begin{equation}
\beta_{\rm QI}(M)
=
\beta_f(M)
\,
\Theta
\left[
{\cal D}_G(k_M)
-
{\cal D}_G^{\rm th}
\right],
\label{eq:beta_qi_simple}
\end{equation}
where $k_M$ is the scale associated with the PBH mass $M$.  This equation
should not be interpreted as a new collapse condition.  It only selects those
PBH-producing modes whose Gaussian discord is above a chosen diagnostic
threshold.

More generally, if the environmental history is stochastic, one may introduce
a distribution over covariance-matrix parameters,
\begin{equation}
\lambda
=
\{r,\eta,n_{\rm env},\varphi\},
\label{eq:lambda_covariance_parameters}
\end{equation}
and write
\begin{equation}
\beta_{\rm QI}(M)
=
\int d\delta_R\,d\lambda\,
P(\delta_R,\lambda;M)
\Theta(\delta_R-\delta_c)
\Theta
\left[
{\cal D}_G(\lambda)
-
{\cal D}_G^{\rm th}
\right].
\label{eq:beta_qi_general}
\end{equation}
This form makes the role of the quantum-information variables explicit, but
it is not yet predictive until the environmental distribution is specified.
For the purposes of the present work, Eq.~\eqref{eq:beta_qi_simple} is
therefore the cleaner diagnostic form. In the pure squeezed limit, the
discord threshold is typically easy to satisfy, since large squeezing already
implies a large Gaussian discord. The more interesting regime is the
decohered mixed state, where entanglement may be lost while weaker quantum
correlations remain. In this regime, Gaussian discord provides a useful
diagnostic of the quantum-to-classical transition of the PBH-producing modes,
without changing the classical collapse condition.

If the open-system dynamics also changes the scalar two-point covariance, then
the scalar power spectrum entering Eq.~\eqref{eq:sigma_delta} should be
replaced by an effective spectrum,
\begin{equation}
{\cal P}_{\zeta}^{\rm eff}(k)
=
Q_{\rm dec}(k)
{\cal P}_{\zeta}^{\rm class}(k).
\label{eq:Pzeta_eff_pbh}
\end{equation}
Here $Q_{\rm dec}(k)$ is a model-dependent covariance-survival factor.  It is
not equal to the discord itself.  In a dissipative attenuation channel,
$Q_{\rm dec}(k)<1$ is possible, while in a pure dephasing channel one may have
$Q_{\rm dec}(k)=1$ even though the density matrix has decohered.

With the replacement in Eq.~\eqref{eq:Pzeta_eff_pbh}, the density variance
becomes
\begin{equation}
\sigma_{\delta,{\rm eff}}^2(R)
=
\int d\ln k\,
W^2(kR)
\left[
\frac{4}{9}
(kR)^2
T(kR)
\right]^2
Q_{\rm dec}(k)
{\cal P}_{\zeta}^{\rm class}(k).
\label{eq:sigma_delta_eff}
\end{equation}
The corresponding PBH abundance is
\begin{equation}
\beta_f^{\rm eff}(M)
=
\frac{1}{2}
{\rm erfc}
\left(
\frac{\delta_c}{\sqrt{2}\sigma_{\delta,{\rm eff}}}
\right).
\label{eq:beta_eff}
\end{equation}
Because $\beta_f$ depends exponentially on $1/\sigma_\delta^2$, even a modest
covariance suppression can reduce $f_{\rm PBH}$ by many orders of magnitude.
For example, if $Q_{\rm dec}$ is approximately constant across a narrow scalar
peak, then $\sigma_{\delta,{\rm eff}}^2\simeq Q_{\rm dec}\sigma_\delta^2$.
A value such as $Q_{\rm dec}=0.7$ therefore changes the exponent in
Eq.~\eqref{eq:beta_asymptotic}, which can strongly suppress the final PBH
fraction.

Thus the role of the quantum-information sector is twofold.  First, Gaussian
discord tracks the residual quantum correlations of the PBH-producing modes.
Second, if the same open-system dynamics also reduces the scalar covariance,
then the PBH abundance is modified through the effective variance in
Eq.~\eqref{eq:sigma_delta_eff}.  These two effects should be kept conceptually
separate.
\section{Scalar-induced gravitational waves and the $Q_{\rm dec}^2$ scaling}
\label{sec:sigw}

\subsection{Standard induced-GW formula}

Scalar perturbations source tensor perturbations at second order in
cosmological perturbation theory.  Working in radiation domination, the tensor
mode with polarization $\lambda$ obeys
\begin{equation}
h_{\bm k,\lambda}''
+
2{\cal H}h_{\bm k,\lambda}'
+
k^2h_{\bm k,\lambda}
=
4S_{\bm k,\lambda},
\label{eq:tensor_equation_sigw}
\end{equation}
where ${\cal H}=a'/a$ is the conformal Hubble parameter and a prime denotes a
derivative with respect to conformal time $\tau$.  The source term
$S_{\bm k,\lambda}$ is quadratic in the scalar perturbations.  Schematically,
it has the form
\begin{equation}
S_{\bm k,\lambda}(\tau)
\sim
\int
\frac{d^3q}{(2\pi)^3}
e^{ij}_{\lambda}(\bm k)
q_iq_j
\zeta_{\bm q}(\tau)
\zeta_{\bm k-\bm q}(\tau),
\label{eq:tensor_source_schematic}
\end{equation}
where $e^{ij}_{\lambda}(\bm k)$ is the transverse-traceless polarization
tensor.  The exact expression contains the radiation-era scalar transfer
functions, but the important point is that the tensor source is quadratic in
the scalar perturbations.

The tensor power spectrum therefore involves a scalar four-point function,
\begin{equation}
\left\langle
\zeta\zeta\zeta\zeta
\right\rangle .
\label{eq:scalar_four_point}
\end{equation}
For Gaussian scalar perturbations, Wick's theorem reduces this four-point
function to products of two scalar power spectra.  Under this assumption, the
present-day scalar-induced gravitational-wave spectrum can be written as
\cite{Ananda:2006af,Baumann:2007zm,Kohri:2018awv,Inomata:2019ivs}
\begin{equation}
\Omega_{{\rm GW},0}(k)
=
c_g\Omega_{r,0}
\frac{1}{24}
\int_0^\infty dv
\int_{|1-v|}^{1+v}du\,
{\cal K}(u,v)
\overline{I^2(u,v)}
{\cal P}_{\zeta}(ku)
{\cal P}_{\zeta}(kv).
\label{eq:sigw_standard}
\end{equation}
Here $\Omega_{r,0}$ is the present radiation density fraction, and
\begin{equation}
c_g
\simeq
0.83
\left(
\frac{g_*}{10.75}
\right)^{-1/3}
\label{eq:cg_factor}
\end{equation}
accounts for the change in the number of relativistic degrees of freedom
between the time of generation and today.  The variables $u$ and $v$ are
defined by
\begin{equation}
u
=
\frac{|\bm k-\bm q|}{k},
\qquad
v
=
\frac{q}{k}.
\label{eq:uv_definitions}
\end{equation}
The geometric kernel is
\begin{equation}
{\cal K}(u,v)
=
\left[
\frac{
4v^2
-
(1+v^2-u^2)^2
}{
4uv
}
\right]^2,
\label{eq:geometric_kernel}
\end{equation}
and $\overline{I^2(u,v)}$ is the time-averaged radiation-era transfer kernel.
The overline denotes averaging over the fast tensor oscillations after the
mode is well inside the horizon.

Equation~\eqref{eq:sigw_standard} is the standard result used in many PBH and
scalar-induced gravitational-wave studies.  It shows that the induced signal
is quadratic in the scalar power spectrum.  This quadratic dependence is the
reason why even a moderate change in the scalar covariance can have a visible
effect on $\Omega_{\rm GW}$.

\subsection{Decohered scalar covariance}

We now connect the open-system description of the scalar sector to the induced
gravitational-wave signal.  We define a covariance-level survival factor
$Q_{\rm dec}(k)$ by
\begin{equation}
Q_{\rm dec}(k)
\equiv
\frac{
{\cal P}_{\zeta}^{\rm eff}(k)
}{
{\cal P}_{\zeta}^{\rm class}(k)
}.
\label{eq:Q_def}
\end{equation}
Equivalently, the equal-time two-point function is written as
\begin{equation}
\left\langle
\zeta_{\bm k}
\zeta_{\bm q}
\right\rangle_{\rm eff}
=
(2\pi)^3
\delta^{(3)}(\bm k+\bm q)
\frac{2\pi^2}{k^3}
Q_{\rm dec}(k)
{\cal P}_{\zeta}^{\rm class}(k).
\label{eq:zeta_two_point_Qdec}
\end{equation}
This definition is important.  The factor $Q_{\rm dec}$ modifies the scalar
power spectrum, or equivalently the scalar two-point covariance.  It is not
the field-amplitude suppression factor.  If one defines an amplitude-level
suppression $Q_{\rm amp}$ by
\begin{equation}
\zeta_{\bm k}^{\rm eff}
=
Q_{\rm amp}(k)
\zeta_{\bm k}^{\rm class},
\label{eq:Qamp_definition}
\end{equation}
then the corresponding power-spectrum factor is
\begin{equation}
Q_{\rm dec}(k)
=
Q_{\rm amp}^2(k).
\label{eq:Qdec_Qamp_relation}
\end{equation}

It should also be stressed that $Q_{\rm dec}$ is not equal to the Gaussian
discord.  Discord diagnoses quantum correlations in the covariance matrix.
The factor $Q_{\rm dec}$ instead parametrizes a possible change in the scalar
two-point covariance caused by a specific open-system channel.  In a pure
dephasing channel, the reduced density matrix can decohere while
$Q_{\rm dec}=1$ for the scalar power spectrum.  In a dissipative attenuation
channel, one can have $Q_{\rm dec}<1$.

Substituting
\begin{equation}
{\cal P}_{\zeta}^{\rm eff}(k)
=
Q_{\rm dec}(k)
{\cal P}_{\zeta}^{\rm class}(k)
\label{eq:P_eff_Qdec_sigw}
\end{equation}
into Eq.~\eqref{eq:sigw_standard}, one obtains
\begin{equation}
\Omega_{{\rm GW},0}^{\rm eff}(k)
=
c_g\Omega_{r,0}
\frac{1}{24}
\int_0^\infty dv
\int_{|1-v|}^{1+v}du\,
{\cal K}(u,v)
\overline{I^2(u,v)}
Q_{\rm dec}(ku)
Q_{\rm dec}(kv)
{\cal P}_{\zeta}^{\rm class}(ku)
{\cal P}_{\zeta}^{\rm class}(kv).
\label{eq:sigw_Q_full}
\end{equation}
This is the full covariance-level result in the static approximation.  The
two factors $Q_{\rm dec}(ku)$ and $Q_{\rm dec}(kv)$ appear because the induced
GW source is quadratic in the scalar perturbations and the tensor power
spectrum is built from products of two scalar power spectra.

If the scalar spectrum is sharply peaked around $k_*$ and
$Q_{\rm dec}(k)$ varies slowly over the support of the convolution, then
\begin{equation}
Q_{\rm dec}(ku)
\simeq
Q_{\rm dec}(kv)
\simeq
Q_{\rm dec}(k_*).
\label{eq:Qdec_slow_variation}
\end{equation}
In this narrow-peak approximation, Eq.~\eqref{eq:sigw_Q_full} reduces to
\begin{equation}
\Omega_{{\rm GW},0}^{\rm eff}(k)
\simeq
Q_{\rm dec}^2(k_*)
\Omega_{{\rm GW},0}^{\rm class}(k).
\label{eq:Q2_scaling}
\end{equation}
This is the origin of the $Q_{\rm dec}^2$ scaling used in the numerical
plots.  The scaling does not mean that quantum discord directly sources
gravitational waves.  It means that, if the scalar two-point covariance is
reduced by a factor $Q_{\rm dec}$, then the induced gravitational-wave power
is reduced by approximately two such factors.

The approximation in Eq.~\eqref{eq:Q2_scaling} can fail if the scalar power
spectrum is broad, if $Q_{\rm dec}(k)$ has rapid scale dependence, or if the
scalar perturbations have significant non-Gaussianity.  In those cases, one
must use the full convolution in Eq.~\eqref{eq:sigw_Q_full}, and possibly also
include the connected scalar four-point function.  In the present work we
restrict ourselves to Gaussian scalar perturbations and use
Eq.~\eqref{eq:sigw_Q_full} as the baseline expression.
\subsection{Time-dependent decoherence}

The discussion above assumed that the decoherence process has already reached
a saturation regime before the scalar modes efficiently source the tensor
perturbations.  In a more general situation, however, the open-system evolution
may continue during radiation domination.  The covariance-survival factor then
depends on both scale and time,
\begin{equation}
Q_{\rm dec}
=
Q_{\rm dec}(k,\tau).
\label{eq:Qdec_time_dependent_sigw}
\end{equation}
In this case the induced gravitational-wave calculation cannot be obtained by
a simple replacement
${\cal P}_{\zeta}(k)\rightarrow Q_{\rm dec}(k){\cal P}_{\zeta}(k)$ with a
time-independent factor.  Instead, the scalar covariance entering the
second-order tensor source should be evaluated at the time when the source is
active.

Schematically, the tensor perturbation is obtained from the Green-function
solution
\begin{equation}
h_{\bm k,\lambda}(\tau)
=
4
\int^{\tau}
d\tau'\,
G_k(\tau,\tau')
S_{\bm k,\lambda}(\tau'),
\label{eq:tensor_green_solution}
\end{equation}
where $G_k(\tau,\tau')$ is the radiation-era tensor Green function and
$S_{\bm k,\lambda}$ is quadratic in scalar perturbations.  Therefore the
tensor power spectrum contains unequal-time scalar correlators.  If the scalar
covariance is affected by open-system dynamics, the time-dependent factors
$Q_{\rm dec}(ku,\tau')$ and $Q_{\rm dec}(kv,\tau'')$ should appear inside the
time integrals that define the radiation-era kernel.

A schematic version of the time-dependent result is
\begin{equation}
\Omega_{{\rm GW},0}^{\rm eff}(k)
\sim
\int dv
\int du
\int d\tau'
\int d\tau''\,
{\cal K}(u,v)\,
{\cal I}(u,v;\tau',\tau'')
Q_{\rm dec}(ku,\tau')
Q_{\rm dec}(kv,\tau'')
{\cal P}_{\zeta}^{\rm class}(ku)
{\cal P}_{\zeta}^{\rm class}(kv),
\label{eq:sigw_time_dependent_schematic}
\end{equation}
where ${\cal I}(u,v;\tau',\tau'')$ denotes the product of scalar transfer
functions and tensor Green functions.  This expression shows why the
time-dependent case is more involved than the static approximation.  A fully
consistent calculation would require solving the open-system evolution of the
scalar covariance together with the second-order tensor equation.

In the present work we restrict ourselves to the simpler saturation regime.
That is, we assume that the relevant decoherence or attenuation process is
completed before, or shortly after, the scalar modes begin to source the
tensor perturbations efficiently.  In this limit one may approximate
\begin{equation}
Q_{\rm dec}(k,\tau)
\simeq
Q_{\rm dec}(k),
\label{eq:Qdec_static_limit_sigw}
\end{equation}
and the induced gravitational-wave spectrum reduces to the static
convolution-level expression in Eq.~\eqref{eq:sigw_Q_full}.  This approximation
is appropriate when the open-system relaxation timescale is shorter than the
Hubble timescale around horizon re-entry.  If this condition is not satisfied,
the suppression can become frequency dependent, and the full time-dependent
treatment should be used.  We leave this more complete analysis for future
work.

\subsection{Relation to discord}

It is important to clarify the relation between $Q_{\rm dec}$ and Gaussian
discord.  The scaling in Eq.~\eqref{eq:Q2_scaling} does not mean that discord
directly sources gravitational waves.  Scalar-induced gravitational waves are
sourced by scalar perturbations through the second-order Einstein equations.
The quantity that enters the standard induced-GW convolution is the scalar
two-point covariance, or equivalently the scalar power spectrum.  Gaussian
discord instead diagnoses the residual quantum correlations contained in the
covariance matrix.

Thus, $Q_{\rm dec}$ and ${\cal D}_G$ have different meanings.  The factor
$Q_{\rm dec}$ parametrizes the survival of the scalar two-point covariance in
a specified open-system channel,
\begin{equation}
Q_{\rm dec}(k)
=
\frac{
{\cal P}_{\zeta}^{\rm eff}(k)
}{
{\cal P}_{\zeta}^{\rm class}(k)
}.
\label{eq:Qdec_power_definition}
\end{equation}
By contrast, ${\cal D}_G$ is a quantum-information measure computed from the
full Gaussian covariance matrix.  A state can lose entanglement and still have
nonzero discord.  It can also decohere in phase without changing the equal-time
power spectrum.  Therefore, there is no universal identity between
$Q_{\rm dec}$ and ${\cal D}_G$.

For a local Gaussian attenuation channel, the scalar power may be written
schematically as
\begin{equation}
{\cal P}_{\zeta}^{\rm eff}(k)
=
\eta(k)
{\cal P}_{\zeta}^{\rm class}(k)
+
\left[
1-\eta(k)
\right]
{\cal P}_{\zeta}^{\rm env}(k),
\label{eq:P_eff_eta_env_sigw}
\end{equation}
where $\eta(k)$ is the transmissivity and
${\cal P}_{\zeta}^{\rm env}(k)$ is the effective environmental contribution.
This gives
\begin{equation}
Q_{\rm dec}(k)
=
\eta(k)
+
\left[
1-\eta(k)
\right]
\frac{
{\cal P}_{\zeta}^{\rm env}(k)
}{
{\cal P}_{\zeta}^{\rm class}(k)
}.
\label{eq:Qdec_eta_env}
\end{equation}
In the cold-environment limit, where
${\cal P}_{\zeta}^{\rm env}(k)\ll{\cal P}_{\zeta}^{\rm class}(k)$, this
reduces to
\begin{equation}
Q_{\rm dec}(k)
\simeq
\eta(k).
\label{eq:Qdec_eta_cold}
\end{equation}
In contrast, for a pure dephasing channel, the density matrix can decohere
while the scalar power spectrum remains unchanged.  In that case one may have
\begin{equation}
Q_{\rm dec}(k)=1,
\label{eq:Qdec_dephasing}
\end{equation}
even though the quantum state has become mixed.

One may introduce a phenomenological diagnostic relation between covariance
survival and discord survival, for example
\begin{equation}
Q_{\rm D}(k)
\equiv
\frac{
{\cal D}_G^{\rm mixed}(k)
}{
{\cal D}_G^{\rm pure}(k)
},
\label{eq:QD_discord_ratio}
\end{equation}
but this quantity should not be automatically identified with
$Q_{\rm dec}(k)$.  The ratio $Q_{\rm D}$ measures the survival of Gaussian
discord, whereas $Q_{\rm dec}$ measures the survival of the scalar power
spectrum.  They may be correlated in a particular microscopic model, but the
relation is model dependent.

The robust result derived in this section is therefore the following:
if the scalar covariance is modified as
\begin{equation}
{\cal P}_{\zeta}^{\rm eff}(k)
=
Q_{\rm dec}(k)
{\cal P}_{\zeta}^{\rm class}(k),
\label{eq:robust_Qdec_statement}
\end{equation}
then the induced gravitational-wave spectrum contains the product
$Q_{\rm dec}(ku)Q_{\rm dec}(kv)$ inside the convolution.  For a narrow scalar
peak and slowly varying $Q_{\rm dec}$, this gives the approximate scaling
\begin{equation}
\Omega_{\rm GW}^{\rm eff}
\simeq
Q_{\rm dec}^2
\Omega_{\rm GW}^{\rm class}.
\label{eq:robust_Qdec2_statement}
\end{equation}
This is a statement about covariance suppression, not a direct sourcing of
gravitational waves by quantum discord.

\section{Numerical Analysis}
\label{sec:numerical}

We now turn to the numerical illustrations.  The purpose of this section is
not to introduce additional assumptions, but to show how the formal results
derived above appear in the figures.  The logic is deliberately sequential.
We first examine the quantum state of the PBH-producing curvature modes, then
follow the effect of decoherence on entanglement and discord, and finally
connect the surviving scalar covariance to PBH abundance and scalar-induced
gravitational waves.  This ordering keeps the interpretation conservative:
PBH formation remains a classical collapse process, while the
quantum-information variables diagnose the state of the perturbations that
seed the collapse and source the induced tensor background.

\begin{figure}[t]
\centering
\includegraphics[width=0.7\columnwidth]{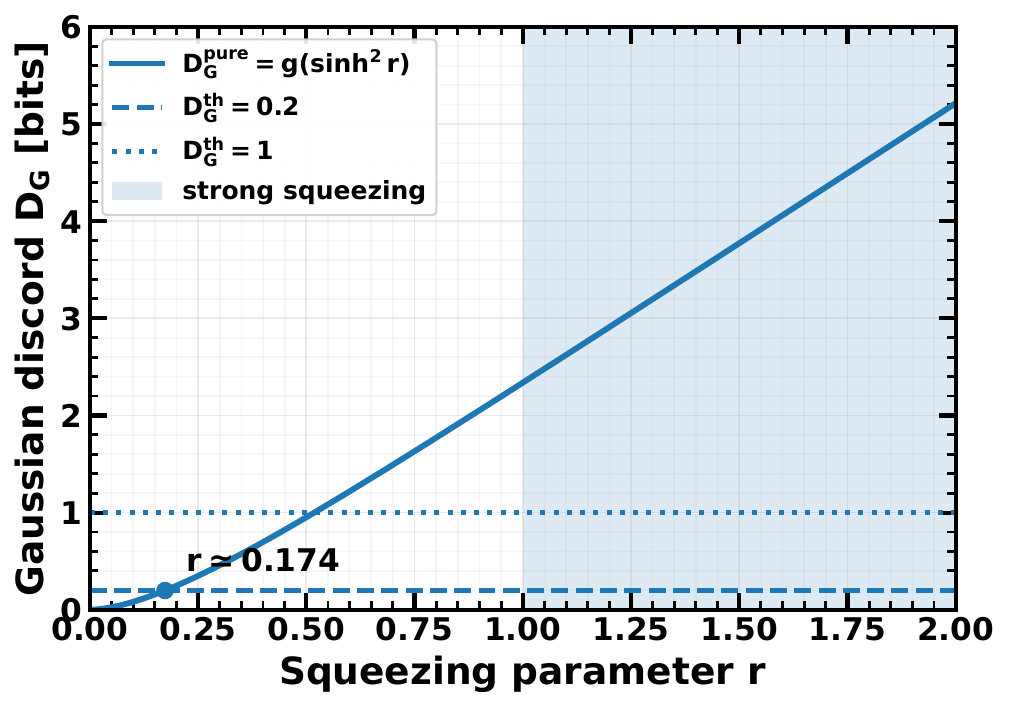}
\caption{
Pure-state Gaussian discord as a function of the squeezing parameter \(r\).
The threshold values \(\mathcal D_G=0.2,1.0\), and \(3.0\) bits are reached
at \(r=0.175,0.519\), and \(1.232\), respectively.  The shaded region denotes
the strongly squeezed regime \(r>1\), where the pure-state discord is already
large.  The figure shows that strongly squeezed PBH-producing modes naturally
carry large discord before decoherence is included.
}
\label{fig:pure_discord}
\end{figure}

Figure~\ref{fig:pure_discord} provides the reference point for the
quantum-information part of the analysis.  It shows the pure-state Gaussian
discord as a function of the squeezing parameter \(r\), using
Eq.~\eqref{eq:pure_discord}.  The numerical crossings shown in the figure are
consistent with Eqs.~\eqref{eq:discord_example_02}--\eqref{eq:discord_threshold_r}.
The thresholds \(\mathcal D_G=0.2,1.0\), and \(3.0\) bits are reached at
\(r=0.175,0.519\), and \(1.232\), respectively.  The shaded region marks
\(r>1\), where the discord is already larger than about \(2.34\) bits.  This
is an important baseline result.  In the absence of decoherence, strong
squeezing automatically implies large Gaussian discord.  Therefore a low
discord threshold is not an independent PBH-formation condition.  It is
instead a diagnostic of the quantum correlations already present in the
two-mode squeezed state.  This interpretation is consistent with the
standard squeezed-state description of inflationary perturbations and with
the use of Gaussian discord as a quantum-correlation measure
\cite{Polarski:1995jg,Kiefer:1998qe,Martin:2015qta,Giorda:2010wpy,
Adesso:2011efy}.

\begin{figure}[t]
\centering
\includegraphics[width=\columnwidth]{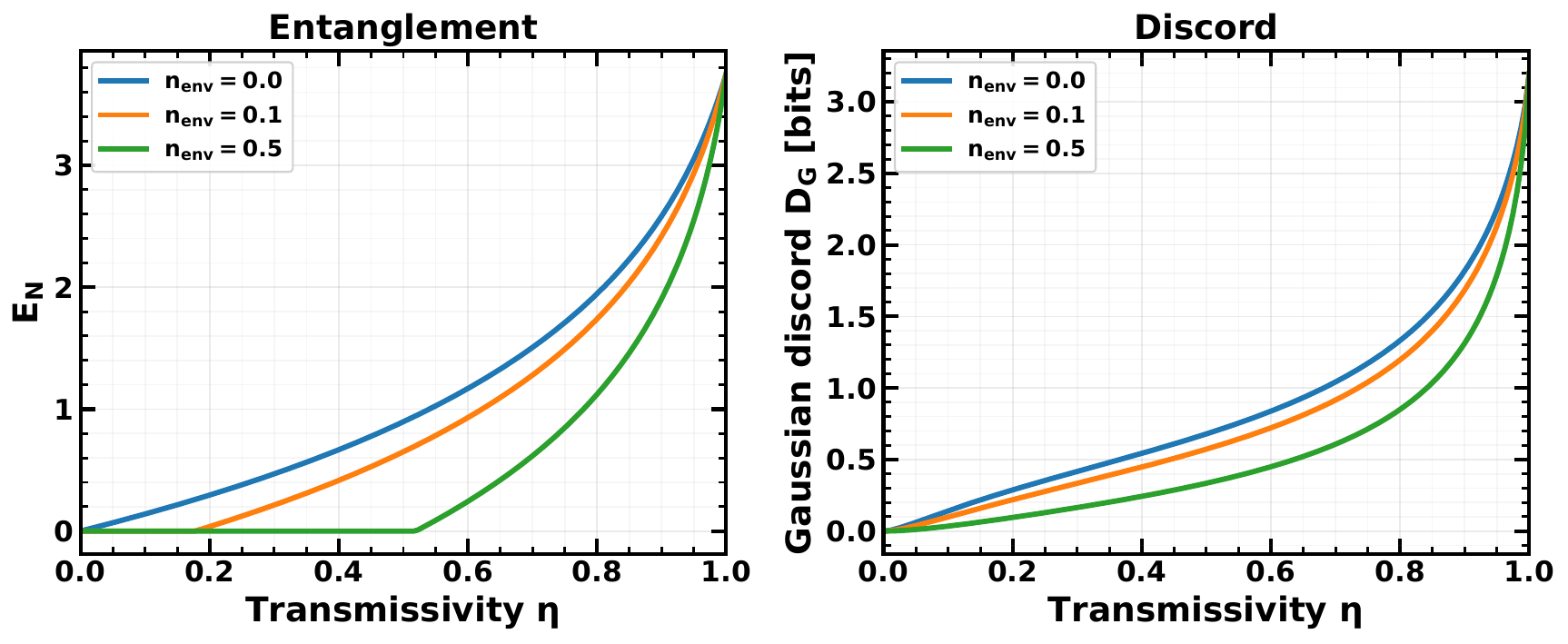}
\caption{
Entanglement and Gaussian discord under the Gaussian decoherence channel.
The logarithmic negativity \(E_N\) is rapidly degraded as the transmissivity
\(\eta\) decreases and can vanish at finite decoherence strength.  Gaussian
discord is more robust and may remain nonzero even after the state becomes
separable.  The figure therefore demonstrates why discord is a useful
diagnostic of residual quantum correlations in the PBH-producing curvature
modes.
}
\label{fig:decoherence_discord}
\end{figure}

The pure-state result is useful, but it is not the main physical regime of
interest.  The more informative case is the mixed state generated by
environmental decoherence.  Figure~\ref{fig:decoherence_discord} illustrates
this regime using the Gaussian attenuation channel described by
Eqs.~\eqref{eq:loss_channel_A}--\eqref{eq:loss_channel_C}.  The left part of
the figure follows the logarithmic negativity, computed from
Eq.~\eqref{eq:log_negativity}.  As the channel transmissivity \(\eta\) is
reduced, the state becomes increasingly mixed and the logarithmic negativity
can vanish at finite decoherence strength.  This is the familiar fragility of
entanglement in an open quantum system.

The right part of Fig.~\ref{fig:decoherence_discord} shows the corresponding
Gaussian discord, computed from Eq.~\eqref{eq:gaussian_discord_mixed}.
Although the discord also decreases as decoherence becomes stronger, it
survives over a larger region of parameter space than the logarithmic
negativity.  The figure therefore separates three physically distinct
regimes: a nearly pure squeezed regime, a mixed entangled regime, and a
separable but still discordant regime.  The last regime is the most relevant
for our purpose.  It shows that the loss of entanglement does not by itself
mean that the curvature perturbations have become a completely classical
stochastic ensemble.  A separable Gaussian state can still retain
phase-space correlations inherited from the original inflationary squeezing
\cite{Giorda:2010wpy,Adesso:2011efy,Weedbrook:2011wxo}.

\begin{figure*}[t]
\centering
\begin{minipage}{0.48\textwidth}
    \centering
    \includegraphics[width=\linewidth]{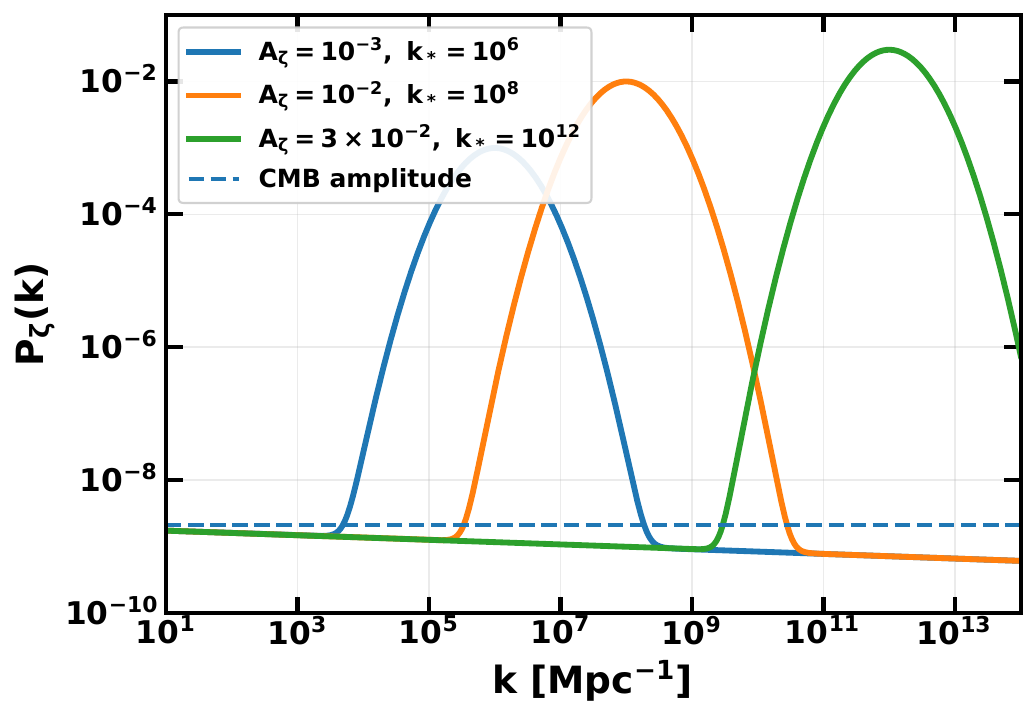}
    \vspace{1mm}
    {\small (a) Curvature spectra and PBH mass scales.}
\end{minipage}
\hfill
\begin{minipage}{0.48\textwidth}
    \centering
    \includegraphics[width=\linewidth]{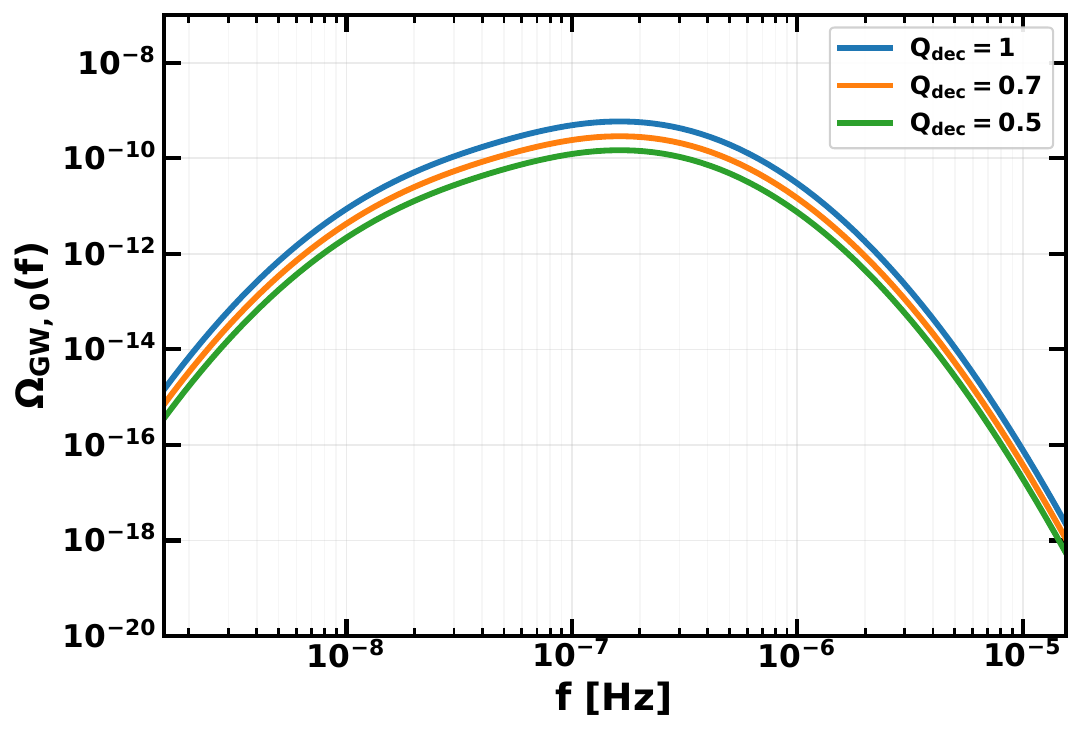}
    \vspace{1mm}
    {\small (b) SIGW spectra with decoherence suppression.}
\end{minipage}
\caption{
Connection between the scalar spectrum, PBH formation, and scalar-induced
gravitational waves.  Panel (a) shows representative peaked curvature spectra.
Changing the peak position shifts the corresponding PBH mass scale, while
changing the amplitude controls the abundance through the classical collapse
probability.  Panel (b) shows the benchmark example of induced gravitational-wave spectra.  In the
constant-\(Q_{\rm dec}\) benchmark, decoherence suppresses the amplitude by
\(Q_{\rm dec}^2\) without changing the spectral shape.  A scale- or
time-dependent decoherence factor would instead lead to a frequency-dependent
distortion.
}
\label{fig:cosmo_summary}
\end{figure*}

We next connect these quantum-information diagnostics to the scalar sector
that controls PBH formation and scalar-induced gravitational waves.  The left
panel of Fig.~\ref{fig:cosmo_summary} shows representative peaked curvature
spectra.  Moving the peak to larger \(k_*\) corresponds to perturbations that
re-enter the Hubble radius earlier, and hence to smaller PBH masses.  Raising
the peak amplitude increases the smoothed density variance in
Eq.~\eqref{eq:sigma_delta}.  Since the classical PBH abundance depends on the
tail of the distribution through Eq.~\eqref{eq:beta}, and becomes
exponentially sensitive in the small-variance limit described by
Eq.~\eqref{eq:beta_asymptotic}, even a modest change in the scalar amplitude
can produce a large change in \(f_{\rm PBH}\).  This is why PBH abundance is
a very sensitive probe of the small-scale scalar spectrum
\cite{Carr:1974nx,Harada:2013epa,Escriva:2019phb,
Musco:2020jjb}.

The same scalar spectra also source induced gravitational waves.  The right
panel of Fig.~\ref{fig:cosmo_summary} shows the corresponding SIGW spectra.
The standard result is controlled by the convolution in
Eq.~\eqref{eq:sigw_standard}.  Once the scalar covariance is modified by a
decoherence factor, the full static expression is Eq.~\eqref{eq:sigw_Q_full}.
This is the equation that should be regarded as the main result for the
decohered SIGW spectrum.  The simpler \(Q_{\rm dec}^2\) scaling shown in the
plot follows only in the narrow-peak and slowly varying \(Q_{\rm dec}\) limit,
as stated in Eq.~\eqref{eq:Q2_scaling}.  Thus, for a constant benchmark
\(Q_{\rm dec}=0.7\), the induced signal is reduced by a factor \(0.49\), while
\(Q_{\rm dec}=0.5\) gives a reduction by \(0.25\).  The shape is unchanged in
this simple limit because the same constant factor multiplies the scalar
covariance across the support of the convolution.  A scale-dependent or
time-dependent decoherence history would instead distort the spectrum and
would require the more general treatment discussed around
Eq.~\eqref{eq:sigw_time_dependent_schematic}.

\begingroup
\color{black}
\subsection{PBH mass fraction and the effect of covariance suppression}

To show explicitly how the covariance-survival factor affects the PBH
abundance, we use Eqs.~\eqref{eq:sigma_delta_eff},
\eqref{eq:beta_eff}, and \eqref{eq:fpbh} together with the mass-scale
relation in Eq.~\eqref{eq:mass_wavenumber_relation}.  The illustrative
calculation assumes a narrow lognormal curvature peak,
\begin{equation}
{\cal P}_{\zeta}^{\rm class}(k)
=
A_\zeta
\exp\left[
-\frac{\ln^2(k/k_*)}{2\sigma_{\ln k}^2}
\right],
\label{eq:lognormal_scalar_peak}
\end{equation}
and a constant covariance-survival factor across the peak,
\begin{equation}
\sigma_{\delta,{\rm eff}}^2(M)
\simeq
Q_{\rm dec}\,\sigma_\delta^2(M).
\label{eq:sigma_eff_constant_Q}
\end{equation}
This approximation is only a benchmark.  A full prediction would require the
model-dependent scalar spectrum, window function, transfer function, and
possible non-Gaussian corrections.  Nevertheless, it cleanly displays the
main effect: because \(\beta_f\) contains
\(\exp[-\delta_c^2/(2\sigma_\delta^2)]\), a moderate reduction of the
variance produces a much larger reduction of \(f_{\rm PBH}\).
\endgroup

\begin{figure}[t]
\centering
\includegraphics[width=0.9\columnwidth]{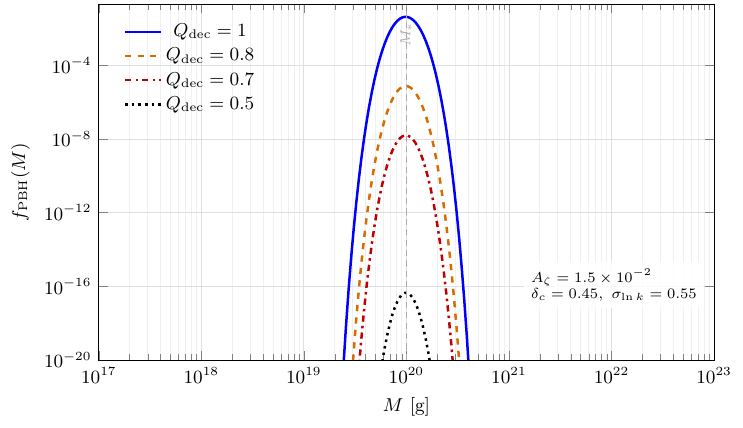}
\caption{
Illustrative PBH dark-matter fraction as a function of PBH mass for constant
covariance-survival factors \(Q_{\rm dec}\).  The benchmark uses
\(A_\zeta=1.5\times10^{-2}\), \(\sigma_{\ln k}=0.55\),
\(\delta_c=0.45\), \(M_*=10^{20}\,{\rm g}\), \(\gamma=0.2\), and
\(g_*=106.75\).  The rapid separation between the curves shows the exponential
sensitivity of PBH production to the effective variance.  Thus decoherence
affects PBH abundance only if the open-system channel suppresses the scalar
two-point covariance, while pure dephasing with \(Q_{\rm dec}=1\) would leave
this abundance unchanged at the power-spectrum level.
}
\label{fig:pbh_fraction_qdec}
\end{figure}

\begingroup
\color{black}
Figure~\ref{fig:pbh_fraction_qdec} shows that the PBH mass distribution keeps
approximately the same mass location when \(Q_{\rm dec}\) is taken constant
over the narrow peak, but its normalization is strongly suppressed.  In the
benchmark shown, reducing the scalar covariance from \(Q_{\rm dec}=1\) to
\(Q_{\rm dec}=0.8\) already lowers the peak fraction by several orders of
magnitude, and \(Q_{\rm dec}=0.7\) suppresses it further.  This behavior is
much stronger than the \(Q_{\rm dec}^2\) scaling of the SIGW spectrum because
PBH formation samples the extreme Gaussian tail, whereas the induced GW signal
is quadratic in the scalar power spectrum.
\endgroup

The four figures should be read as a single chain of results.  Figure
\ref{fig:pure_discord} shows that the PBH-producing modes naturally carry
large pure-state discord once they are strongly squeezed.  Figure
\ref{fig:decoherence_discord} shows that the mixed-state problem is
nontrivial: entanglement can be erased while discord remains nonzero.  Figure
\ref{fig:cosmo_summary} \and Fig.~\ref{fig:pbh_fraction_qdec} then show where the same covariance information can
enter cosmological observables.  The PBH abundance is affected only if the
open-system channel changes the classical scalar variance entering
Eq.~\eqref{eq:sigma_delta_eff} and hence the effective collapse fraction in
Eq.~\eqref{eq:beta_eff}.  By contrast, the SIGW signal is directly sensitive
to the scalar covariance inside Eq.~\eqref{eq:sigw_Q_full}.  This is the
main reason why induced gravitational waves provide a particularly clean
place to look for decoherence-suppressed scalar correlations.

The numerical examples therefore support a conservative interpretation.  The
standard PBH/SIGW calculation uses the scalar power spectrum, the classical
collapse threshold, and the induced-GW kernel.  The quantum-information
extension keeps the same collapse criterion, but follows the full covariance
matrix and distinguishes pure squeezed, mixed entangled, and separable but
discordant regimes.  Finally, a specific dissipative decoherence model may
introduce a covariance-survival factor \(Q_{\rm dec}\).  This factor modifies
the induced gravitational-wave spectrum according to Eq.~\eqref{eq:sigw_Q_full}
and reduces to the simple scaling in Eq.~\eqref{eq:Q2_scaling} only under the
narrow-peak approximation.  Its effect on PBH abundance is more
model-dependent, because PBH formation depends on the real-space density
variance and the nonlinear collapse threshold.  In this sense, quantum discord
is not a new source of PBHs or gravitational waves.  Its value is that it
tracks the quantum-to-classical transition of the scalar perturbations whose
covariance later appears in PBH and SIGW observables.

\section{Conclusions}
\label{sec:conclusions}

We have studied how decoherence of PBH-producing curvature perturbations can
affect primordial black-hole abundance and the associated scalar-induced
gravitational-wave background. The PBH collapse criterion was kept classical:
PBHs form when the density perturbation exceeds the threshold at horizon
re-entry. Gaussian quantum discord was used only as a diagnostic of residual
quantum correlations in the squeezed scalar modes, not as a new condition for
PBH formation.

Using a Lindblad-inspired Gaussian loss channel, we separated two different
effects. The first is the survival of quantum correlations, measured by
Gaussian discord. The second is the possible suppression of the scalar
two-point covariance, parametrized by \(Q_{\rm dec}(k)\). Only the second
effect directly changes PBH and SIGW observables. For PBHs, it enters through
the smoothed density variance and can strongly affect the abundance because
the collapse probability is exponentially sensitive to this variance. For
scalar-induced gravitational waves, the effect is more direct because the
induced tensor source is quadratic in scalar perturbations. Therefore, in the
narrow-peak limit and for slowly varying \(Q_{\rm dec}\), the spectrum scales
approximately as \(Q_{\rm dec}^{2}\Omega_{\rm GW}^{\rm class}\).

The central result is that decoherence and residual quantum correlations can
be incorporated into the standard PBH/SIGW framework without changing the
classical collapse picture. In this formulation, Gaussian discord characterizes
the quantum-to-classical transition of the scalar sector, while
\(Q_{\rm dec}\) controls the observable covariance-level effect. This provides
a consistent way to relate the quantum history of PBH-producing perturbations
to measurable signatures in scalar-induced gravitational waves.

\bibliographystyle{unsrt}
\bibliography{ref}
\end{document}